\documentstyle[aps,multicol,tighten,epsfig,array]{revtex}

\begin{document}

\draft

\title{Ponderomotive entangling of atomic motions}

\author{Stefano Mancini$^1$ and
Sougato Bose$^2$}

\address{
$^1$INFM, Dipartimento di Fisica,
Universit\`a di Milano,
Via Celoria 16, I-20133 Milano, Italy
\\
$^2$Centre for Quantum Computation, Clarendon Laboratory,
University of Oxford, Parks Road,
Oxford OX1 3PU, England}

\date{\today}

\maketitle

\begin{abstract}
We propose the use of ponderomotive forces to entangle the motions
of different atoms. Two situations are analyzed: one where the
atoms belong to the same optical cavity and interact with the
same radiation field mode; the other where each atom is placed in
own optical cavity and the output field of one cavity enters the
other.
\end{abstract}


\begin{multicols}{2}

\section{Introduction}

The preparation of entangled atomic states is one of the goals of
atomic physics and quantum optics. These states are the key
ingredients for studying some fundamental issues of quantum
mechanics \cite{BELL}, as well as  for certain applications
related to quantum information \cite{BEDI}. Various methods have
been recently proposed to engineer entanglement between atoms
\cite{HAR,ATOMS,PAR}. They are based on achieving and controlling
an effective interaction between the atoms that are to be
entangled. Typically, these interactions are mediated by the
electromagnetic field, but also involve transitions between
internal atomic states.

On the other hand, atoms and radiation fields can interact via
radiation pressure effects. The role of such {\it ponderomotive
effects} in probing fundamental aspects of quantum theory has been
pointed out in a number of recent papers \cite{OUR1,OUR2}. Here
we shall exploit ponderomotive effects to propose a method of
entangling atomic motions. This is fundamentally important as
radiation pressure effects are universal and scalable. For
example, radiation pressure effects could also be fruitful in
entangling massive particles (or even macroscopic objects). If
such a scheme happens to be successful for atoms, one would
acquire the confidence of trying out a similar scheme with more
massive objects. Another advantage of a ponderomotive scheme
becomes clear when we note that most existing mechanisms for
entangling atoms, apart from a very few \cite{PAR}, entangle
internal atomic states. As such, the maximum degree of
entanglement attainable is limited by the finite dimensionality
of the Hilbert space of the internal states of the interacting
atoms. Our scheme, on the other hand, entangles atomic motions
and therefore entangles two continuous variable (infinite
dimensional) systems. Moreover, as we shall demonstrate, our
entangling mechanism, in contrast to most others, does not
require carefully controlled switching on and off of external
laser fields acting on the individual atoms.

As Hamiltonian model, we shall consider the case of large
detuning of internal atomic transitions from the cavity field so
that spontaneous emission can be neglected and the upper atomic
level can be adiabatically eliminated \cite{WALMIL}. In this case,
the atom-field interaction reduces to the product between the
number of photons and the amplitude of atomic displacement. The
atomic internal states are never involved in the interaction and
the atom can always stay in a fixed internal ground state.

\section{Atoms in the same optical cavity}

We first consider two trapped atoms $A$ and $B$
in the same optical cavity. When they are invested
by the (off-resonant) laser light, the evolution
in one spatial direction takes
place according to the Hamiltonian (in natural units)
\begin{equation}\label{H1}
H=\chi c^{\dag}c \left(a+a^{\dagger}+b+b^{\dagger}\right) +\Omega
(a^{\dagger}a + b^{\dagger} b) \,
\end{equation}
where $a,b$ and $c$ are annihilation operators for the vibrational
motion of the two atoms and the cavity mode respectively.
Furthermore, $\Omega$ is the vibrational frequency (assumed
equal for the two atoms).
The time evolution operator corresponding to the Hamiltonian
(\ref{H1}) can be put in the following form \cite{OUR1}
\begin{eqnarray}\label{U}
U(t)&=&\exp\left[2i\kappa^2\left(c^{\dag}c\right)^2
\left(t-\sin t\right)\right]
\nonumber\\
&&\times\exp\left[\kappa c^{\dag}c
\left(\eta(t)a^{\dag}-\eta^*(t)a\right)\right]
\nonumber\\
&&\times\exp\left[\kappa c^{\dag}c
\left(\eta(t)b^{\dag}-\eta^*(t)b\right)\right]
\nonumber\\
&&\times\exp\left[-it\left(a^{\dag}a+b^{\dag}b\right)\right]
\,,
\end{eqnarray}
where $\eta(t)=(1-e^{-it})$, $\kappa=\chi/\Omega$
and the time is scaled accordingly to $\Omega t\to t$.

Let us now assume that initially both atoms are  cooled down to their
ground states
and the cavity field is in coherent states, that is
\begin{equation}\label{Psiini}
|\Psi(0)\rangle= |0\rangle_a
\otimes |0\rangle_b 
\otimes|\zeta\rangle_c\,,
\end{equation}
then, the time evolution leads to
\begin{eqnarray}\label{Psit}
|\Psi(t)\rangle&=&e^{-|\zeta|^2/2}
\sum_{n=0}^{\infty}\frac{\zeta^n}{\sqrt{n!}}
e^{2i\kappa^2 n^2(t-\sin t)}
\nonumber\\
&&|n\kappa\eta(t)\rangle_a
\otimes |n\kappa\eta(t)\rangle_b
\otimes|n\rangle_c\,,
\end{eqnarray}
where $|n\kappa\eta(t)\rangle_{a,b}$ are coherent states
of the atoms.

Let us suppose to measure the quadrature $X=(c+c^{\dag})/\sqrt{2}$.
Then the state after the measurement will be
\begin{equation}\label{Proj}
|{\widetilde\Psi}(t)\rangle={\cal N}\,
|x\rangle_c {}_c\langle x|\Psi(t)\rangle\,,
\end{equation}
where ${\cal N}$ is a normalization constant,
while $|x\rangle$ are the eigenvectors of the
quadrature observable $X$.
The inverse of normalization constant also gives the probability
amplitude of the outcome $x$, that is $P(x)={\cal N}^{-2}$.

The joint state of the atoms after the measurement results
\begin{eqnarray}\label{Psitil}
|\widetilde{\Psi}(t)\rangle&=&{\cal N}e^{-|\zeta|^2/2}
\sum_{n=0}^{\infty}
\frac{\zeta^n}{\sqrt{n!}}e^{2i\kappa^2n^2(t-\sin t)}
{}_c\langle x|n\rangle_c
\nonumber\\
&&|n\kappa\eta(t)\rangle_a
\otimes |n\kappa\eta(t)\rangle_b\,,
\end{eqnarray}
where
\begin{equation}\label{xn}
{}_c\langle x|n\rangle_c=\pi^{-1/4}(2^n n!)^{-1/2}H_n(x)\exp(-x^2/2)\,,
\end{equation}
are the harmonic oscillator position eigenstates,
with $H_n$ the Hermite polynomials.

The state (\ref{Psitil}) depends on the time at which
the measurement is performed and is conditioned
to the result $x$ of the measurement.
Let us choose a time $t_N=(2N+1)\pi$ with $N\in{\bf N}$,
so that $\eta(t_N)=2$,
and $\kappa\in{\bf N}$, thus
Eq.(\ref{Psitil}) becomes
\begin{equation}\label{PsitN}
|\widetilde{\Psi}(t_N)\rangle={\cal N}
e^{-|\zeta|^2/2}\sum_{n=0}^{\infty}
\frac{\zeta^{n}}{\sqrt{n!}}
\,{}_c\langle x|n\rangle_c\,
|4\kappa n\rangle_a \otimes |4\kappa n\rangle_b\,.
\end{equation}
It is clear that the above state represent an entangled state
for the atoms.
In practice, since the radiation field mediates information between
the two atoms,
a measurement of its quadrature leaves the atoms correlated.

\begin{figure}[t]
\centerline{\epsfig{figure=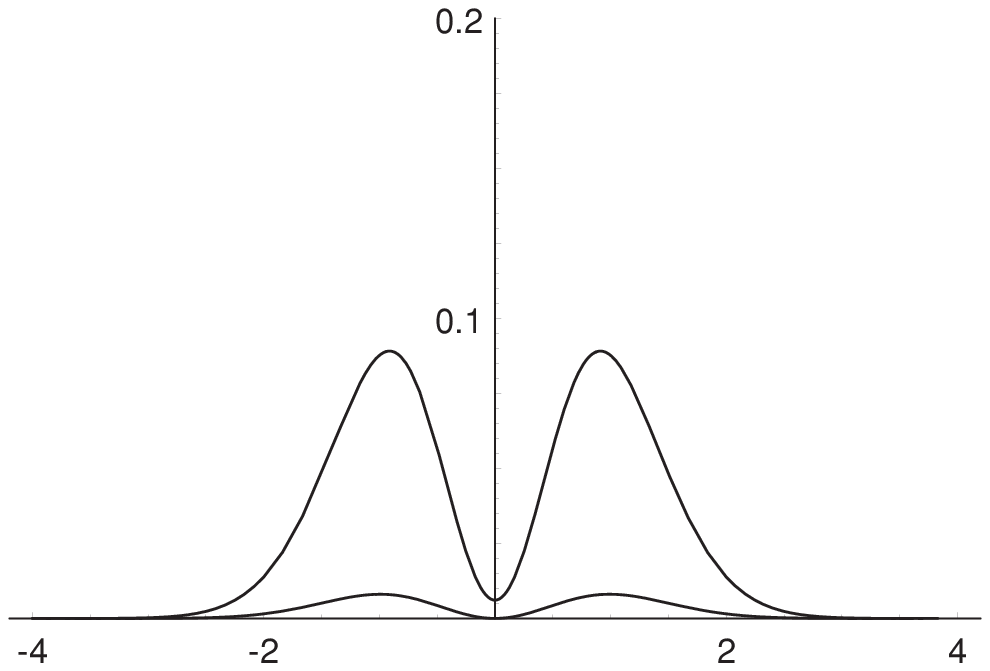,width=3in}}
\caption{\narrowtext
Entanglement efficiency vs quadrature outcome for different
values of $\zeta$ (from bottom to top, $\zeta=0.1$, $0.4$, $0.8$).
Here, $\kappa=1$.}
\label{fig1}
\end{figure}

In order to quantify the degree of entanglement we calculate the
linear entropy \cite{JEX}
\begin{equation}\label{Elin}
E=1-{\rm Tr}_a\{[{\rm Tr}_b(\,\widetilde\rho\,)]^2\}\,,
\end{equation}
where $\widetilde\rho=
|\widetilde\Psi(t_N)\rangle \langle\widetilde\Psi(t_N)|$.
However, since the formation of entangled states is conditioned
to the measurement on the radiation field, it would be
useful to define the efficiency of the
entanglement procedure as
\begin{equation}\label{Ups}
\Upsilon(x)=E(x)\times P(x)\,.
\end{equation}
Then, in Fig.\ref{fig1} we show the efficiency $\Upsilon$
as function of quadrature outcome $x$. We clearly see that
the efficiency increases as the radiation pressure (i.e.,
the amplitude of the radiation field) increases.
The shape of $\Upsilon$
comes from the fact that $x=0$ is the
most probable outcome of the measurement,
but the entanglement has a minimum at that value.

Note that by increasing $\kappa$ sufficiently, the set of states
$\{|4\kappa n\rangle\}$ in Eq.(\ref{PsitN}) can be made to
approach an orthonormal basis arbitrarily close.  This means that
we can approach the maximal entanglement possible in any $D\times
D$ system arbitrarily closely by setting the field amplitude
$\zeta$ to a required value and increasing $\kappa$. For example,
for a $10\times 10$ system, the maximum entanglement according to
the measure of Eq.(\ref{Elin}) is $0.9$. This is already exceeded
for a field amplitude $\zeta=6$. The above fact clearly
illustrates one of the advantages of entangling through our
scheme in contrast to entangling the internal levels of two
two-level atoms as in most existing schemes. The degree of
entanglement achievable is not bounded from above by any
fundamental constraint. Of course, the degree of entanglement one
can practically produce will depend on parameters such as the
Q-factor of the cavity in a specific experimental realization.

\section{Atoms in distinct optical cavities}

We now consider the two (trapped) atoms placed in separate cavities,
and interacting sequentially with the same radiation field.
That is, the outgoing field from the first cavity enters the second
one.
Therefore, the Hamiltonian (\ref{H1}) should be
modified as follows:
\begin{eqnarray}\label{H2}
H&=&\chi c_1^{\dag}c_1(a+a^{\dag})
+\chi  c_2^{\dag}c_2(b+b^{\dag})
\nonumber\\
&&+\Omega(a^{\dag}a+b^{\dag}b)
\,,
\end{eqnarray}
where we have assumed the same coupling constant and the same
oscillatory frequency for the two atoms.

However, in such a case we have to consider photon losses,
which we assume to occour at same rate $\gamma$ in the two
cavities.
We further assume
a decay of the atomic motions at rate $\Gamma$.
Thus, we can write down the quantum Langevin equations as
\begin{eqnarray}
{\dot a}&=&-i\Omega a-i\chi c_1^{\dag}c_1-\frac{\Gamma}{2}a
+\sqrt{\Gamma}a^{in}\,,
\label{QLE1}
\\
{\dot b}&=&-i\Omega b-i\chi c_2^{\dag}c_2-\frac{\Gamma}{2}b
+\sqrt{\Gamma}b^{in}\,,
\label{QLE2}
\\
{\dot c_1}&=&-i\Delta_1 c_1-i\chi c_1(a+a^{\dag})
-\frac{\gamma}{2}c_1+\sqrt{\gamma}c_1^{in}\,,
\label{QLE3}
\\
{\dot c_2}&=&-i\Delta_2 c_2-i\chi c_2(b+b^{\dag})
-\frac{\gamma}{2}c_2+\sqrt{\gamma}c_2^{in}\,,
\label{QLE4}
\end{eqnarray}
where all the input operators represent vacuum noise \cite{GAR}.
$\Delta_1$ and $\Delta_2$ are the cavity detunings.
The boundary condition reads
\begin{equation}\label{BOU}
c_2^{in}\equiv c_1^{out}=\sqrt{\gamma}c_1-c_1^{in}\,.
\end{equation}

To solve the system of Eqs.(\ref{QLE1}), (\ref{QLE2}),
(\ref{QLE3}), (\ref{QLE4}),
we proceede by the linearization around the steady state.
The latter is characterized by
\begin{eqnarray} \label{zeta12}
&&\zeta_{j}\left\{\frac{\gamma}{2}+\frac{\chi^2|\zeta_{j}|^2\Gamma}
{\frac{\Gamma^2}{4}+\Omega^2}\right.
\nonumber\\
&&\qquad\left.
+i\left[\Delta_{j}-2\frac{\chi^2|\zeta_{j}|^2\Omega}{\frac{\Gamma^2}{4}
+\Omega^2}\right]\right\}
=\sqrt{\gamma}\zeta_{j}^{in}\,;
j=1,2\,,
\end{eqnarray}
with the relation
\begin{equation}\label{zetabou}
\zeta_2^{in}\equiv\zeta_1^{out}=\sqrt{\gamma}\zeta_1-\zeta_1^{in}\,.
\end{equation}
Here, $\zeta$ is the steady state value of the operators $c$,
and $\zeta_1^{in}$ is the amplitude of the input field
(at the first cavity).
Eq.(\ref{zeta12}) shows a typical bistable behavior \cite{MT94}.
Furthermore, the steady states of atomic operators
$a$ and $b$ are given by
\begin{eqnarray}
\alpha&=&-i\frac{\chi}{\frac{\Gamma}{2}+i\Omega}|\zeta_1|^2\,,
\label{al}
\\
\beta&=&-i\frac{\chi}{\frac{\Gamma}{2}+i\Omega}|\zeta_2|^2\,.
\label{be}
\end{eqnarray}

The linearized system of equations can be written in the
frequency domain as
\begin{equation}\label{matrixeq}
i\omega \tilde{v}={\cal M}\tilde{v}+\tilde{v}_{in}
\end{equation}
where the transposed vectors $\tilde{v}^{T}$ and $\tilde{v}_{in}^{T}$
are given by
\begin{equation}\label{vvec}
\tilde{v}^{T}=(
\tilde{a},
\tilde{a}^{\dag},
\tilde{b},
\tilde{b}^{\dag},
\tilde{c}_1,
\tilde{c}_1^{\dag},
\tilde{c}_2,
\tilde{c}_2^{\dag})\,,
\end{equation}
and
\begin{eqnarray}\label{vinvec}
\tilde{v}_{in}^{T}&=&(
\sqrt{\Gamma}\tilde{a}_{in},
\sqrt{\Gamma}\tilde{a}_{in}^{\dag},
\sqrt{\Gamma}\tilde{b}_{in},
\sqrt{\Gamma}\tilde{b}_{in}^{\dag},
\nonumber\\
&&\sqrt{\gamma}\tilde{c}^{in}_1,
\sqrt{\gamma}\tilde{c}_1^{in\,\dag},
-\sqrt{\gamma}\tilde{c}^{in}_2,
-\sqrt{\gamma}\tilde{c}_2^{in\,\dag})\,,
\end{eqnarray}
where now all the operators represent small quantum fluctuations
around steady state.
Moreover, the $8\times 8$ matrix ${\cal M}$ is
\begin{equation}\label{calM}
{\cal M}=\left[
\begin{array}{cc}
M_{I}&M_{II}
\\
M_{III}&M_{IV}
\end{array}
\right]
\end{equation}
with

\end{multicols}

\begin{onecolumn}

\begin{equation}\label{MI}
M_{I}=\left[
\begin{array}{cccc}
-\left(\frac{\Gamma}{2}+i\Omega\right)&0&0&0
\\
0&-\left(\frac{\Gamma}{2}-i\Omega\right)&0&0
\\
0&0&-\left(\frac{\Gamma}{2}+i\Omega\right)&0
\\
0&0&0&-\left(\frac{\Gamma}{2}-i\Omega\right)
\end{array}
\right]\,,
\end{equation}

\begin{equation}\label{MIIMIII}
M_{II}=i\chi\left[
\begin{array}{cccc}
-\zeta_1^*&-\zeta_1&0&0
\\
\zeta_1^*&\zeta_1&0&0
\\
0&0&-\zeta_2^*&-\zeta_2
\\
0&0&\zeta_2^*&\zeta_2
\end{array}
\right]\,;
\quad
M_{III}=i\chi\left[
\begin{array}{cccc}
-\zeta_1&-\zeta_1&0&0
\\
\zeta_1^*&\zeta_1^*&0&0
\\
0&0&-\zeta_2&-\zeta_2
\\
0&0&\zeta_2^*&\zeta_2^*
\end{array}
\right]\,,
\end{equation}

\begin{equation}\label{MIV}
M_{IV}=\left[
\begin{array}{cccc}
-\frac{\gamma}{2}
-i\left[\Delta_1+\chi\left(\alpha+\alpha^*\right)\right]&0&0&0
\\
0&-\frac{\gamma}{2}
+i\left[\Delta_1+\chi\left(\alpha+\alpha^*\right)\right]&0&0
\\
\gamma&0&-\frac{\gamma}{2}
-i\left[\Delta_2+\chi\left(\beta+\beta^*\right)\right]&0
\\
0&\gamma&0&-\frac{\gamma}{2}
+i\left[\Delta_2+\chi\left(\beta+\beta^*\right)\right]
\end{array}
\right]\,.
\end{equation}

\end{onecolumn}

\begin{multicols}{2}

The solution of the Eq.(\ref{matrixeq}) can formally  be written as
\begin{equation}\label{matrixsol}
\tilde{v}(\omega)=
\left[i\omega{\cal I}-{\cal M}\right]^{-1} \tilde{v}_{in}(\omega)\,,
\end{equation}
where ${\cal I}$ is the $8\times 8$ identity matrix.
Then, the various frequency correlations can be easily calculated
by using the correlations of the vacuum input noise \cite{GAR}.
These should deserve to quantify the entanglement of atomic motions.
Nevertheless, since we deal with non pure states, it is very
difficult to quantify the
degree of entanglement \cite{PLENIO}.
To reach this goal we shall proceed as follows.

We first introduce the dimensionless
atomic position and momentum variable
\begin{eqnarray}\label{qpab}
q_{a}&=&(a+a^{\dag})\,,q_{b}=(b+b^{\dag})\,,
\\
p_{a}&=&-i(a-a^{\dag})\,,p_{b}=-i(b-b^{\dag})\,,
\end{eqnarray}
Now, if the atoms are entangled, one could {\it infer} position
or momentum of one atom through position or momentum of the other
\cite{REID}. The errors of these inferences are then quantified
by the variances $\langle (q_a+q_b)^2 \rangle$ and $\langle
(p_a-p_b)^2 \rangle$. Once the product of these inference errors
lies below the limit of the Heisenberg principle, i.e. $\langle
(q_a+q_b)^2 \rangle\dot\langle (p_a-p_b)^2 \rangle \le
|\langle[q_a,p_a]\rangle|^2/4$, an EPR-like paradox arises
\cite{EPR}. This is a typical manifestation of the existence of
purely quantum correlations between the two systems \cite{REID}.
It is known that when $\langle (q_a+q_b)^2 \rangle\dot\langle
(p_a-p_b)^2 \rangle$ is less than $1$ \cite{cont}, the state is
entangled {\em irrespective} of whether it is pure or mixed.
Though the criteria for the presence of entanglement in continuous
variable systems is present \cite{cont,simon}, and we use this to
prove the presence of entanglement in our case, there is as yet no
rigorously proved measure of continuous variable entanglement. We
shall use a quantity motivated from the above separability
criterion to evaluate the degree of entanglement.

Now, given an operator $\tilde{\cal O}(\omega)$
in the frequency domain,
we define the hermitian operator
${\cal R}_{\{\tilde{\cal O}\}}(\omega)=
[\tilde{\cal O}(\omega)+\tilde{\cal O}(-\omega)]/2$.
Then, recalling the previous argument,
we can define the degree of entanglement as
\begin{equation}\label{Epseudo}
E(\omega)=\frac{
\langle {\cal R}^2_{\{\tilde{q}_{a}+\tilde{q}_{b}\}}(\omega) \rangle
\,
\langle {\cal R}^2_{\{\tilde{p}_{a}-\tilde{p}_{b}\}}(\omega) \rangle}
{\frac{1}{4} \left|
\langle \left[{\cal R}_{\tilde{q}_a}(\omega),
{\cal R}_{\tilde{p}_a}(\omega)\right]
\rangle\right|^2}\,,
\end{equation}
which can be considered as a signature of
entanglement whenever it goes below one.
Notice that this condition is much stronger
than the simple entanglement requirement,
in that $E<1$ requires EPR-type correlations.

\begin{figure}[t]
\centerline{\epsfig{figure=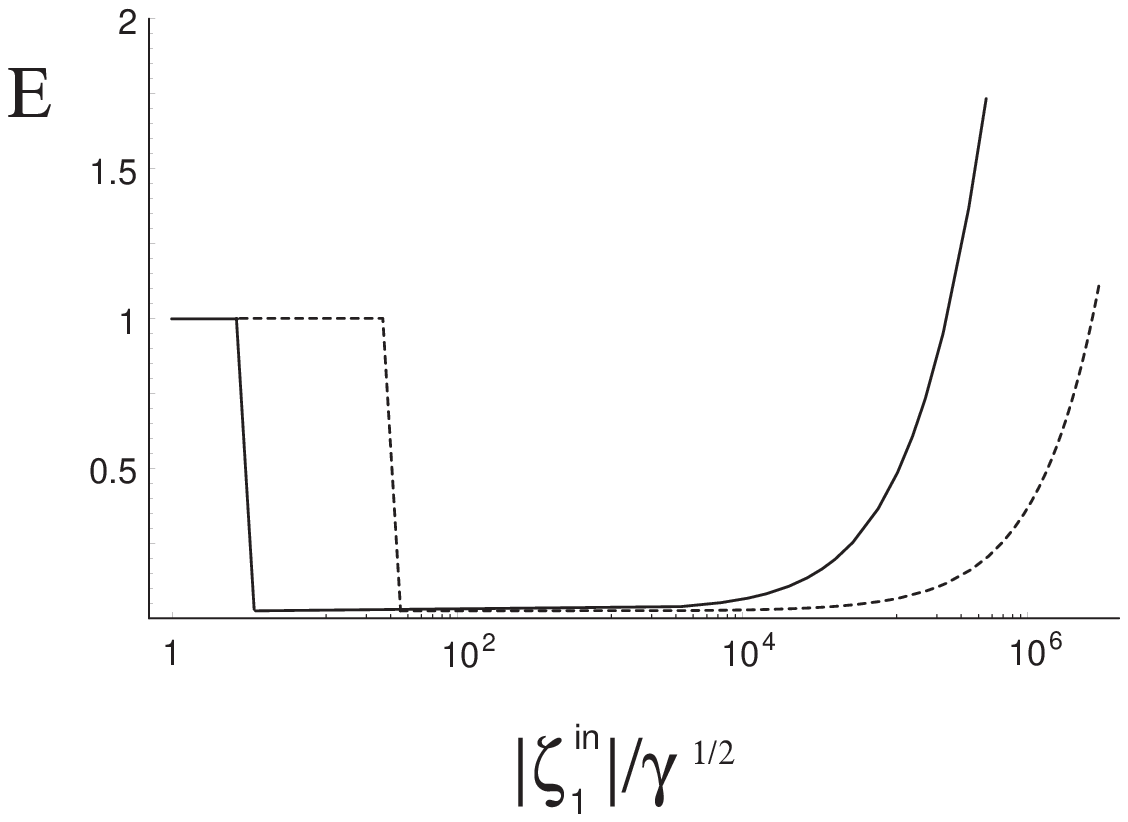,width=3in}}
\caption{\narrowtext
Degree of entanglement vs input field amplitude at frequency $\Omega$.
The values of parameters (in unit of $\gamma$) are:
$\chi=1$ (solid line), $\chi=0.1$ (dashed line) and
$\Delta_1=\Delta_2=10^4$, $\Gamma=10^{-3}$.
}
\label{fig2}
\end{figure}

In Fig.\ref{fig2} we show the degree of entanglement
(\ref{Epseudo}) as function of the input field amplitude. The
sharp decrement is due to the jump from one to the other branch
of the bistable curve \cite{MT94}. Then, by increasing the
radiation intensity, the entanglement tends to disappear because
of the increasing radiation pressure noise. In case of small
coupling constant, entanglement effect appears at higher
amplitudes (dashed line). Notice that we have used nonzero
detunings to establish correlations between $q$ and $p$ variables,
while naturally only $q$ variables tend to couple as results from
Hamiltonian (\ref{H2}).

\section{Conclusions}

In this paper we have discussed how to exploit radiation pressure
effects to entangle the motions of two atoms. We have considered
two scenarios: atoms in same and in separate cavities. As
radiation pressure effects are very generic, the scheme should
lead to those for more macroscopic objects. It also has the
advantage of not involving either atomic internal states or
various laser pulses being applied to the atoms. It is precisely
this fact that offers the generality of our scheme, in the sense
that it should not depend on the specific internal configuration
of the atoms. Moreover, the entanglement between atoms in separate
cavities is generated in the steady state ({\em i.e.} after all
types of decoherence and dissipation have acted). This means that
this type entanglement generating mechanism is robust in nature.
Further generalizations of this scheme for entangling several
atoms interacting with a common cavity field would be interesting
and could potentially provide a simple way for generating
multiparticle Schroedinger cat states.

\section*{Acknowledgements}
S. M. gratefully acknowledges financial support from
Universit\`a di Camerino, Italy, under the Project
`Giovani Ricercatori'.

\end{multicols}


\begin{references}

\bibitem{BELL}
J. S. Bell, Physics {\bf 1}, 195 (1965);
J. F. Clauser, M. A. Horne, A. Shimony and R. A. Holt,
Phys. Rev. Lett. {\bf 23}, 880 (1969).

\bibitem{BEDI}
C. H. Bennett, Phys. Today {\bf 48}(10), 24 (1995);
D. P. DiVincenzo, Science {\bf 270}, 255 (1995).

\bibitem{HAR}
E. Hagley, X. Mantre, G. Nogues, C. Wunderlich, M. Brune,
J. M. Raimond and S. Haroche, Phys. Rev. Lett. {\bf 79}, 1 (1997); 
C. A. Sackett, D. Kielpinski, B. E. King, C. Langer, 
V. Meyer, C. J. Myatt, M. Rowe, Q. A. Turchette, W. M. Itano 
and D. J. Wineland, Nature {\bf 404}, 256 (2000).

\bibitem{ATOMS}
J. I. Cirac and P. Zoller, Phys. Rev. A {\bf 50}, R2799 (1994);
T. Pellizzari, S. Gardiner, J. I. Cirac and P. Zoller, Phys. Rev.
Lett. {\bf 75}, 3788 (1995); J. I. Cirac and P. Zoller, Phys.
Rev. Lett. {\bf 74}, 4091 (1995); J. F. Poyatos, J. I. Cirac and
P. Zoller, Phys. Rev. Lett. {\bf 81}, 1322 (1998); C. Cabrillo,
J. I. Cirac, P. Garcia-Fernandez and P. Zoller, Phys. Rev. A {\bf
59}, 1025 (1999);  M.B. Plenio, S.F. Huelga, A. Beige and P.L.
Knight, Phys. Rev. A {\bf 59} (1999) 2468; S. Bose, P. L. Knight,
M. B. Plenio and V. Vedral, Phys. Rev. Lett. {\bf 83}, 5158
(1999); A. Kuzmich and E. S. Polzik, Phys. Rev. Lett. {\bf 85},
5639 (2000); A. Beige, W. J. Munro and P. L. Knight, Phys. Rev. A
{\bf 62} 2102, (2000); S. Bose and D. Home, quant-ph/0101093.

\bibitem{PAR}
A. S. Parkins, J. Opt. B.: Quant. Semiclass. Opt. {\bf 3},
S18 (2001); A. S. Parkins and H. J. Kimble, Phys. Rev. A {\bf
61}, 052104 (2000).

\bibitem{OUR1}
S. Mancini, V. I. Man'ko and P. Tombesi,
Phys. Rev. A {\bf 55}, 3042 (1997);
S. Bose, K. Jacobs and P. Knight, Phys. Rev. A {\bf 56},
4175 (1997).

\bibitem{OUR2}
S. Bose, K. Jacobs and P. Knight, Phys. Rev. A {\bf 59}, 3204
(1999); S. Mancini, D. Vitali and P. Tombesi,  Phys. Rev. Lett.
{\bf 80}, 688 (1998); S. Mancini, Phys. Lett. A {\bf 279}, 1
(2001); S. Mancini and A. Gatti,
J. Opt. B.: Quant. Semiclass. Opt. {\bf 3},
S66 (2001);
V. Giovannetti,
S. Mancini and P. Tombesi, Europhys. Lett. (to appear).

\bibitem{WALMIL}
D. F. Walls and G. J. Milburn, {\it Quantum Optics}, (Springer,
Berlin, 1994), p.330.

\bibitem{JEX}
G. Drobny, I. Jex and V. Buzek, 
Phys. Rev. A {\bf 48}, 569 (1993).

\bibitem{GAR}
Gardiner, C. W.
{\it Quantum Noise},
(Springer, Berlin, 1991).

\bibitem{MT94}
S. Mancini and P. Tombesi,
Phys. Rev. A {\bf 49}, 4055 (1994).

\bibitem{PLENIO}
V. Vedral and M. B. Plenio,
Phys. Rev. A {\bf 57}, 1619 (1998);
S. Parker, S. Bose and M. B. Plenio,
Phys. Rev. A {\bf 61}, 032305 (2000).

\bibitem{REID}
M. D. Reid and P. D. Drummond,
Phys. Rev. Lett. {\bf 60}, 2731 (1988);
M. Reid,
Phys. Rev. A {\bf 40}, 913 (1989).

\bibitem{EPR}
A. Einstein, B. Podolsky and N. Rosen,
Phys. Rev. {\bf 47}, 777 (1935).

\bibitem{cont}
Lu-Ming Duan, G. Giedke, J. I. Cirac and P. Zoller, Phys. Rev.
Lett. {\bf 84}, 2722 (2000).

\bibitem{simon}
R. Simon, Phys. Rev. Lett. 84, 2726 (2000).

\end{references}
\end{document}